\documentclass[aps,showpacs,prl,twocolumn,superscriptaddress]{revtex4}
\usepackage[pdftex]{graphicx}
\usepackage{amssymb}
\usepackage{amsmath}
\usepackage{textcomp}
\usepackage{times}
\begin{document}

\date{\today}

\title{Quantitative characterization of the nanoscale local lattice strain induced by Sr dopants in La$_{1.92}$Sr$_{0.08}$CuO$_4$}

\author{J. Q. Lin}
\affiliation{Beijing National Laboratory for Condensed Matter Physics and Institute of Physics, Chinese Academy of Sciences, Beijing 100190, China}
\affiliation{School of Physical Sciences, University of Chinese Academy of Sciences, Beijing 100049, China}

\author{X. Liu}
\email{liuxr@shanghaitech.edu.cn} 
\affiliation{Beijing National Laboratory for Condensed Matter Physics and Institute of Physics, Chinese Academy of Sciences, Beijing 100190, China}
\affiliation{School of Physical Science and Technology, ShanghaiTech University, Shanghai 201210, China}
\affiliation{Collaborative Innovation Center of Quantum Matter, Beijing, China}

\author{E. Blackburn}
\affiliation{School of Physics and Astronomy, University of Birmingham, Birmingham B15 2TT, UK}

\author{S. Wakimoto}
\affiliation{Materials Sciences Research Center, Japan Atomic Energy Agency, Tokai, Ibaraki 319-1195, Japan}

\author{H. Ding}
\affiliation{Beijing National Laboratory for Condensed Matter Physics and Institute of Physics, Chinese Academy of Sciences, Beijing 100190, China}
\affiliation{Collaborative Innovation Center of Quantum Matter, Beijing, China}

\author{Z. Islam}
\affiliation{Advanced Photon Source, Argonne National Laboratory, Argonne, IL 60439 USA}

\author{S. K. Sinha}
\affiliation{Department of Physics, University of California, San Diego, La Jolla, California 92093, USA}

\begin{abstract}
The nanometer scale lattice deformation brought about by the dopants in high temperature superconducting cuprate La$_{2-x}$Sr$_x$CuO$_4$(x=0.08) was investigated by measuring the associated X-ray diffuse scattering around multiple Bragg peaks. A characteristic diffuse scattering pattern was observed, which can be well described by continuum elastic theory. With the fitted dipole force parameters, the acoustic type lattice deformation pattern was re-constructed and found to be of similar size to lattice thermal vibration at 7 K. Our results address the long-term concern of dopant introduced local lattice inhomogeneity, and show that the associated nanometer scale lattice deformation is marginal and cannot, alone, be responsible for the patched variation in the spectral gaps observed with scanning tunneling microscopy in the cuprates. 
\end{abstract}


\maketitle
Many transition metal oxides with strong electron correlations are electronically inhomogeneous\cite{Dagotto2005}, including the superconducting cuprates\cite{Cren2000,pan2001,Lang2002,Kinoda2003,Kato1, Kato2, NMR}. The scanning tunneling microscopy (STM) studies on different cuprate families revealed remarkable nanoscale variations in their essential electron spectral features, including the pseudogap\cite{pan2001,McElroy,Hoffman2012} and the superconducting gap\cite{Gomes2007, Parker2010}. Although earlier theoretical work proposed spontaneous electronic phase separation\cite{Emery1997}, experimental evidences established the correlation between the observed nanoscale electronic variations and the distribution of the dopants\cite{McElroy, Hoffman2012, Hoffman2014}. In addition to contributing carriers to the CuO$_2$ planes, the dopants inherently introduce disorders to the systems which perturb the local electronic structure. Such perturbation to the electronic structure from the disorder is more recently discussed in the context of charge density wave formation in the cuprates\cite{CDW1, CDW2}. In general, a dopant can introduce perturbations of a few kinds: a local dopant potential, intra-unitcell atomic distortion, and the associated nanometer scale strain\cite{McElroy}. Which is the leading local variable in creating the nanoscale electronic inhomogeneity remains an issue to be explored.     

Most of the theoretical work\cite{StaticPotential01,StaticPotential02,StaticPotential03,StaticPotential04,StaticPotential05} has been focused on the local potentials associated with the dopant atoms. With either phenomenologically assigned screening lengths or just single site impurity potentials, the nanoscale variations in carrier density and pairing strength can be reproduced. In the meantime, the local impact of the dopants to the lattice was also explored with extended X-ray absorption fine structure (EXAFS) technique\cite{EXAFS1, EXAFS2, EXAFS3}, which revealed significant atomic deformation near the dopants. On the other hand, such deformation happens within just a few angstroms. It remains unclear how to relate such very local deformation to the nanoscale electronic inhomogeneity observed. In addition to the very strong deformation of the atoms nearest to the dopants, such lattice response will propagate away from the dopant center to form long range strain. The impact of strain on the superconductivity in the cuprates has long been noticed by pressure experiments\cite{hydrostatic} where T$_c$ was tuned on the order of 1K/GPa, and showed strong anisotropic behaviors\cite{uniaxial}. Thus, the dopant-induced local lattice strain, depending on its strength, potentially can be the source of the observed inhomogeneity.

As a perturbation to the ideal periodic lattice, the local strain extended from the dopants manifests itself as diffuse tails from the Bragg points in the X-ray scattering measurements, known as  ``Huang diffuse scattering''(HDS)\cite{HDS0,HDS}. Although most of the STM work has been performed on Bi$_2$Sr$_2$CaCu$_2$O$_{8+\delta}$, this cuprate family is not suitable for HDS measurements because its reciprocal space is overwhelmed by the superstructure modulation satellites\cite{SM2212}, leaving separation of the scattering signal from the strain difficult. Instead, underdoped La$_{2-x}$Sr$_x$CuO$_4$ with a relatively simpler structure was chosen, where the nanoscale electronic inhomogeneities were also reported\cite{Kato1, Kato2}. Clear HDS patterns due to the lattice strain associated with the randomly distributed Sr dopants were observed. Interestingly, the scattering strength of the lattice deformation from such strain is comparable to the thermal diffuse scattering at the low temperature of 7 K, indicating that the size of the dopant induced strain is small. From the reconstructed strain pattern based on continuum elastic theory\cite{HDS}, we conclude that the lattice deformation is of the order of 0.001$\AA$. Although extended to nanometer scale, such small strain is unlikely to be the leading local variable in creating the startling nanoscale electronic inhomogeneity observed with STM.

Underdoped La$_{2-x}$Sr$_x$CuO$_4$(x=0.08) crystals were grown by the traveling-solvent floating-zone technique. This doping level is specifically chosen such that the samples have the minimum doping to be in the superconducting phase, but largely avoiding overlapping of the strain field associate with individual dopants. Single crystals were oriented, cut, and polished to a thickness of 0.4 mm with a facet of 2 X 4 mm$^2$. To remove the stress formed during sample preparation, the samples were annealed in oxygen flow (825 \textcelsius\; for 24 h and then 500 \textcelsius\; for 20 h  with one atm pressure). From SQUID magnetization measurements, T$_c$ was found to be 21.1K with the transition width $\Delta T$ to be less than 1K, indicating high sample quality. X-ray Scattering studies were performed on the 4-ID-D beamline at the Advanced Photon Source (APS). A Si(111) double-crystal monochromator was used to select 20keV x-rays. A pair of Pd-coated mirrors was used to deliver a focused beam on the sample. NaI scintillator was used as a point detector. The crystal axes $(a,b,c)$ and the reciprocal space Miller indices $[H K L]$ (given in the reciprocal lattice units (r.l.u.)) are defined with the low temperature orthorhombic structure. Data presented was collected at 7 K unless mentioned otherwise.

\begin{figure}[t]
\includegraphics[width=0.48\textwidth]{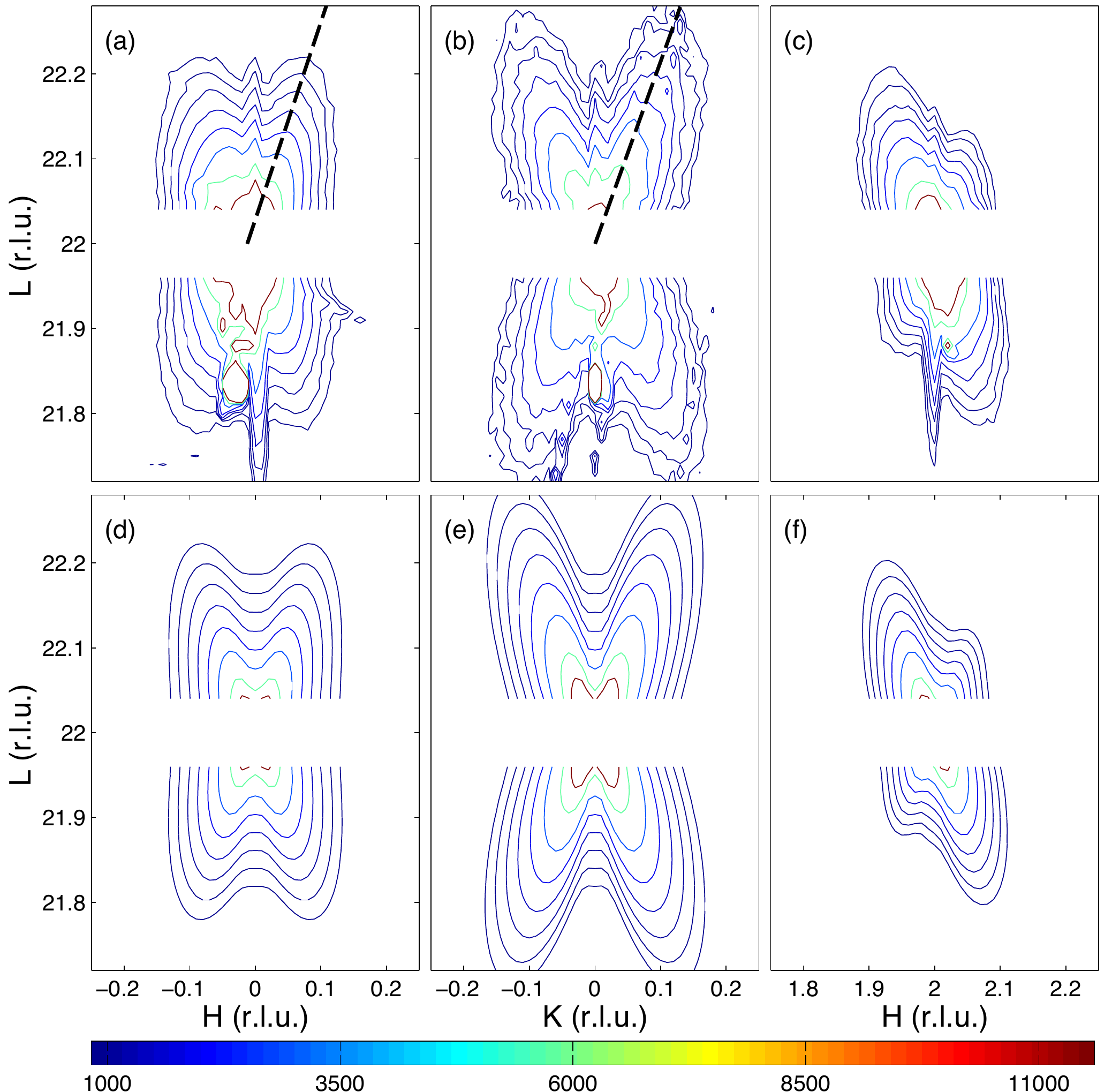}
\caption{Diffuse scattering pattern around (0 0 22) and (2 0 22) Bragg peaks. a-c: experimental data in the {\it H-L (K=0)} and {\it K-L (H=0)} planes, a-b are around (0 0 22) and c is around (2 0 22). Diagonal dashed lines are along the ridges of HDS. d-f: calculated diffuse scattering pattern with both HDS and TDS contributions. The parameters to generate HDS are from our fitting results. Data were taken at $\sim7$ K.}
\label{fig:HKLmesh}
\end{figure}

La$_{1.92}$Sr$_{0.08}$CuO$_4$ undergoes a high temperature tetragonal to low temperature orthorhombic structure phase transition at $\sim280$ K\cite{HTTLTO}, and forms twinned domains at low temperature. Indeed, Bragg peak splitting due to the twinned domains was observed on our sample with large X-ray beam spot size. By narrowing down the X-ray beam spot size to be about 0.1X0.2 $mm^2$ and translating the sample, we managed to perform the measurements on a single domain crystal grain with negligible direction confusion about {\bf a}-{\bf b} axes and the domain averaging effect. As a result, the observed diffuse scattering pattern shown in Fig.~\ref{fig:HKLmesh} is highly anisotropic along $H$ and $K$ directions, reflecting the low temperature orthorhombic {\it Bmab} symmetry.

Fig.~\ref{fig:HKLmesh}(a-c) show the X-ray diffuse scattering intensity mappings for different planes in the reciprocal space around different Bragg peaks. Unlike the thermal diffuse scattering (TDS) which is expected to be more rounded\cite{TDS01}, the observed diffuse scattering pattern show pronounced ridges and valleys, characteristic for HDS. To quantitatively analyze the measured X-ray diffuse scattering patterns, we follow the theory developed by M. A. Krivoglaz and P. H. Dederichs\cite{HDS} for X-ray diffuse scattering from local lattice strains associated with randomly distributed point defects. As the elastic continuum approximation is employed, the details within a unit cell are ignored, and the deformation is in the form of unit cell mass center motion. The HDS around the Bragg point, $\vec{G}$, can be written as,
\begin{equation}
I_{HDS}(\vec{Q})=Nc(1-c)|F(\vec{G})|^2(\frac{Q}{q})^2|\frac{1}{V_c}\sum_{i,j,l}{\hat{Q}_i\tilde{g}_{ij}P_{jl}}\hat{q}_l|^2
\label{eqn:HDS}
\end{equation}
where N is the number of unit cells and $c$ is the doping concentration. $F(\vec{G})$ is the unit cell structure factor at the Bragg point $\vec{G}$. $\vec{q}$ is defined as $\vec{q}=\vec{Q}-\vec{G}$, the reduced vector from Bragg center. $\hat{Q}_i$ and $\hat{q}_l$ are the components of the unit vectors along $\vec{Q}$ and $\vec{q}$ directions. $\tilde{g}_{ij}$ is the inverse of the tensor $\sum_{k,l}{C_{ikjl}q_kq_l/q^2}$ with $C_{ikjl}$ to be the elastic constant. The impact from the defect to its surrounding lattice is contained in the $P_{jl}$ matrix named as ``dipole tensor''\cite{HDS}, which essentially governs the overall HDS pattern. 

The diffuse scattering around the Bragg point $\vec{G}$ also contains contribution from lattice thermal vibrations\cite{TDS}. With acoustic approximation, the TDS can be written as,
\begin{equation}
I_{TDS}(Q)=\frac{N\hbar}{2M}|F(\vec{G})|^2\sum_{j}{\frac{[\vec{Q}\cdot\tilde{e}_{j}(\vec{q})]^2}{\omega_{j}(\vec{q})}}\coth(\frac{\hbar\omega_{j}(\vec{q})}{2K_{B}T})
\label{eqn:TDS}
\end{equation} 
where $M$ is the unit cell mass and $j$ is the index of acoustic branches. $\tilde{e}_{j}(\vec{q})$ and $\omega_{j}(\vec{q})$ are the eigenvectors and eigenfrequencies of the acoustic phonon modes, evaluated from the elastic dynamical matrix.

\begin{figure}[h]
\includegraphics[width=0.48\textwidth]{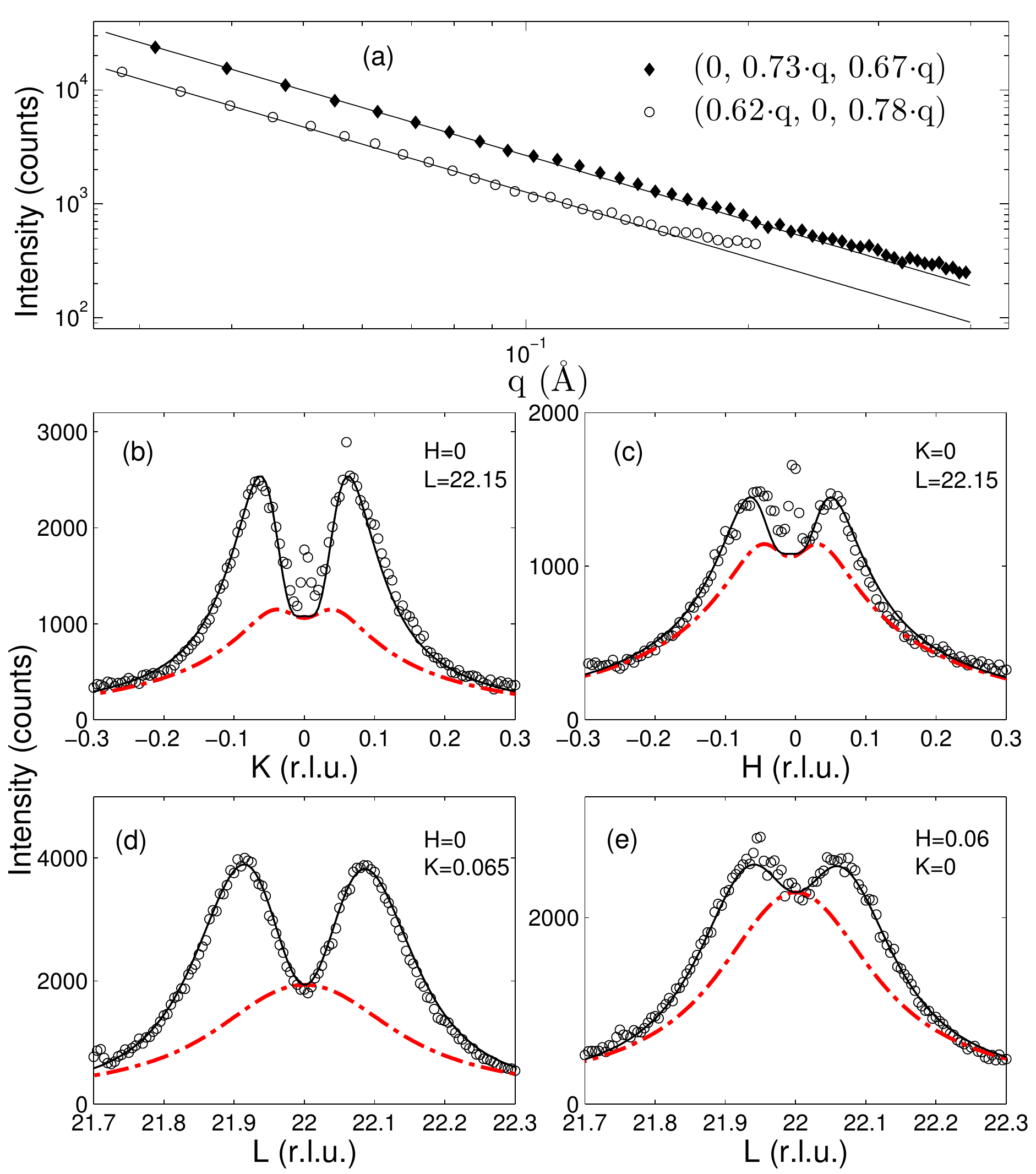}
\caption{(a): diagonal cuts from (0 0 22) peak on the ridges of the diffuse scattering pattern as shown in Fig.~\ref{fig:HKLmesh}(a) and (b). Solid lines are exponential of $q$ as $1/q^{1.9}$. (b)-(e): line cuts near (0 0 22) Bragg peak. Solid lines are the overall fitting results, and the dashed red lines are the TDS components. The sharp central peaks in (b) and (c) are from the crystal truncation rods due to c-direction terminated surface.}
\label{pic:qRelation}
\end{figure}

For a fixed direction of the reduced vector from the Bragg center, Eqn.(\ref{eqn:HDS}) shows that the HDS follows a general $|q|$ dependence as $1/q^\nu$ with $\nu=2$. For TDS, the $|q|$ dependence is temperature dependent. $\nu$ is close to 2 at high temperature and approaches 1 when $\frac{\hbar\omega_{j}(\vec{q})}{2K_{B}T}\gg 1$ at low temperature. To examine the $|q|$ dependence of our diffuse scattering intensity, diagonal cuts in the {\it H-L (K=0)} and {\it K-L (H=0)} planes from (0 0 22) center Bragg peak on the intensity ridges (shown as dashed lines in Fig.~\ref{fig:HKLmesh}(a-b)) were taken and shown in log-log scale in Fig.(\ref{pic:qRelation}). At low $|q|$ region, the intensity shows nicely a $1/q^\nu$ dependence with $\nu=1.9$. The small deviation from 2 is expected due to TDS contributions. In the large $|q|$ region where $\frac{\hbar\omega_{j}(\vec{q})}{2K_{B}T}\gg 1$ is satisfied, the diffuse scattering intensity starts to bend up towards $\nu=1$ as the TDS becomes dominating. Such $|q|$ dependence agrees well with the predictions from Eqn.(\ref{eqn:HDS}) and (\ref{eqn:TDS}), and justify the elastic continuum approximation treatment.  

With both HDS and TDS contributions included, we fitted the total diffuse scattering around different Bragg peaks with the $P_{jl}$ dipole tensor as the fitting parameter. The existence of TDS actually helps our quantitative analysis in the sense that it serves as a built-in reference to the HDS. Since the TDS is completely determined by the temperature and the dynamic matrix which has been already determined\cite{Dmatrix}, the $P_{jl}$ can be determined on an absolute scale. The dipole force $P_{jl}$ respects the site symmetry of the dopants. In the case of Sr in La$_{2-x}$Sr$_x$CuO$_4$, the off-diagonal term $P_{jl}$ with $j$ or $l = 1$ is strictly zero. There are small $P_{23}$ and $P_{32}$ due to the buckling of the CuO$_6$ octahedra, which is ignored in our fitting. As a result, the combination of HDS and TDS formulated in Eqn.(1) and (2) with three diagonal $P_{jj}$ as fitting parameters well reproduces the experimental observations, as shown in Fig.~\ref{fig:HKLmesh}(d-f) and Fig.~\ref{pic:qRelation}(b-e). The $P_{jl}$ dipole tensor is determined as,
\[P = \left( \begin{array}{ccc}
5.23\pm{0.06} & 0 & 0 \\
0 & 8.96\pm{0.06} & 0 \\
0 & 0 & 0.62\pm{0.03} \end{array} \right)10^{-19}N\cdot m\] 
The significant difference between $P_{11}$ and $P_{22}$ reflects the observed strong anisotropy of the diffuse scattering in the {\it H-L (K=0)} and {\it K-L (H=0)} planes around (0 0 22) Bragg peak, suggesting that the lattice around the Sr defects are more strongly distorted along {\bf b} direction than along {\bf a} direction. This anisotropy is a natural consequence of the orthorhombic lattice environment in which the Sr defects reside in. We emphasize that, at this point, the overall sign to the $P_{jl}$ matrix is arbitrary since it appears as the modular square in Eqn.(\ref{eqn:HDS}).  

\begin{figure}[h]
\includegraphics{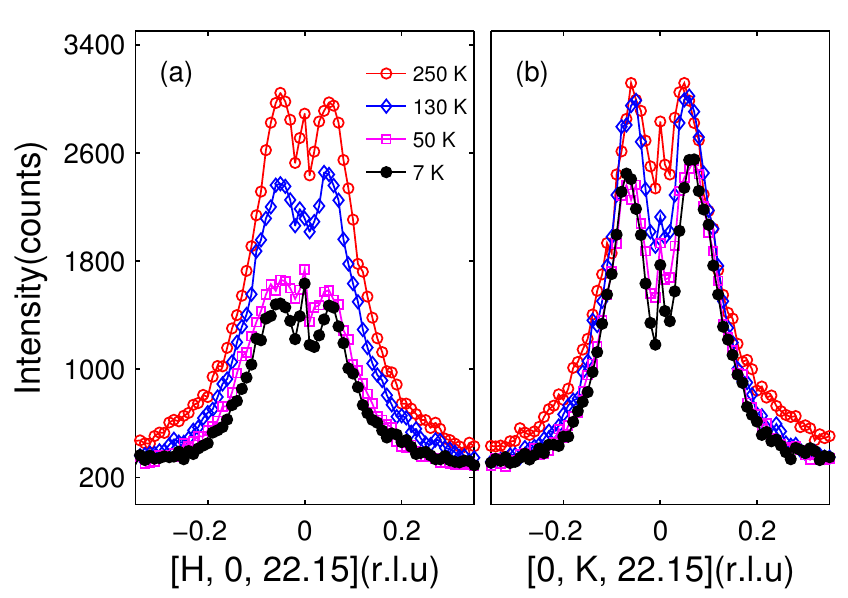}
\caption{Temperature dependent $H$ and $K$ scans across [0, 0, 22.15]. Low temperature data shows big intensity ratio between K and H cuts. The ratio decreases as temperature is increased and almost approaches 1 at $T = 250$ K.}
\label{pic:Tdependent}
\end{figure}

Indeed, our temperature dependent studies further reveal the sensitivity of the dopant associated strain to the subtle lattice symmetry evolutions of La$_{2-x}$Sr$_x$CuO$_4$. In Fig.(\ref{pic:Tdependent}), the line cuts of [$H$, 0, 22.15] and [0, $K$, 22.15] are plotted as a function of temperature. At low temperatures, the strong anisotropy in the diffuse scattering intensities along {\bf a}$^*$ and {\bf b}$^*$ directions can be clearly seen. As a function of increasing temperature, the anisotropy becomes smaller and almost invisible at $T=250$ K, which is close to the HTT-LTO transition temperature($\sim280$ K\cite{HTTLTO}) for $8\%$ doping. The peak-dip-peak feature, characteristic for HDS, gradually loses its strength as the temperature is increased for the [0, $K$, 22.15] cuts shown in Fig.(\ref{pic:Tdependent})(b). Such behavior suggests that the local strain is gradually relaxed as the unit cell volume thermally expands at high temperature, which supports the choice of positive sign to the $P_{jl}$ matrix and the strain is of tensile character. This also agrees with the fact that the ionic radius of Sr$^{2+}$ is much larger than that of La$^{3+}$, leading to an expansion of the unit cell where the dopant resides in.     

With the determined dipole force tensor $P_{jl}$ on absolute scale, the amount of lattice distortion introduced by Sr dopants can be in principle calculated. The displacement from average position at $\vec{R}$ away from the dopant center, $U(\vec{R})$, can be written as\cite{HDS},
\begin{equation}
U(\vec{R})=\frac{i}{(2\pi)^3}\int_{q}\frac{1}{q^2}\sum_{jl}{\tilde{g}_{ij}P_{jl}q_{l}}e^{i\vec{q}\cdot\vec{R}}\mathrm{d} q^3
\label{eqn:displacement}
\end{equation}    

Obviously, this integral diverges as the integration range of $q$ increases. This comes as no surprise since the elastic continuum approximation is only proper at not too large $q$ region. An ellipsoid cut-off was made to the integration in the reciprocal space\cite{SM}. The maximum $q_a$, $q_b$ and $q_c$ are chosen to be 0.25, 0.25 and 0.35 in {\it{r.l.u.}} respectively. These cut-off values were chosen based on the acoustic phonon dispersion curve reported in ref.~\cite{cutoff}, beyond which the dispersion significantly deviates from linear relation and the elastic continuum approximation becomes invalid. We emphasize that these cut-offs are also consistent with our data. As shown in Fig.~\ref{pic:qRelation}, beyond these values the measured diffuse scattering intensity starts to merge into the background and becomes indistinguishable. 

\begin{figure}[h]
\includegraphics[width=0.48\textwidth]{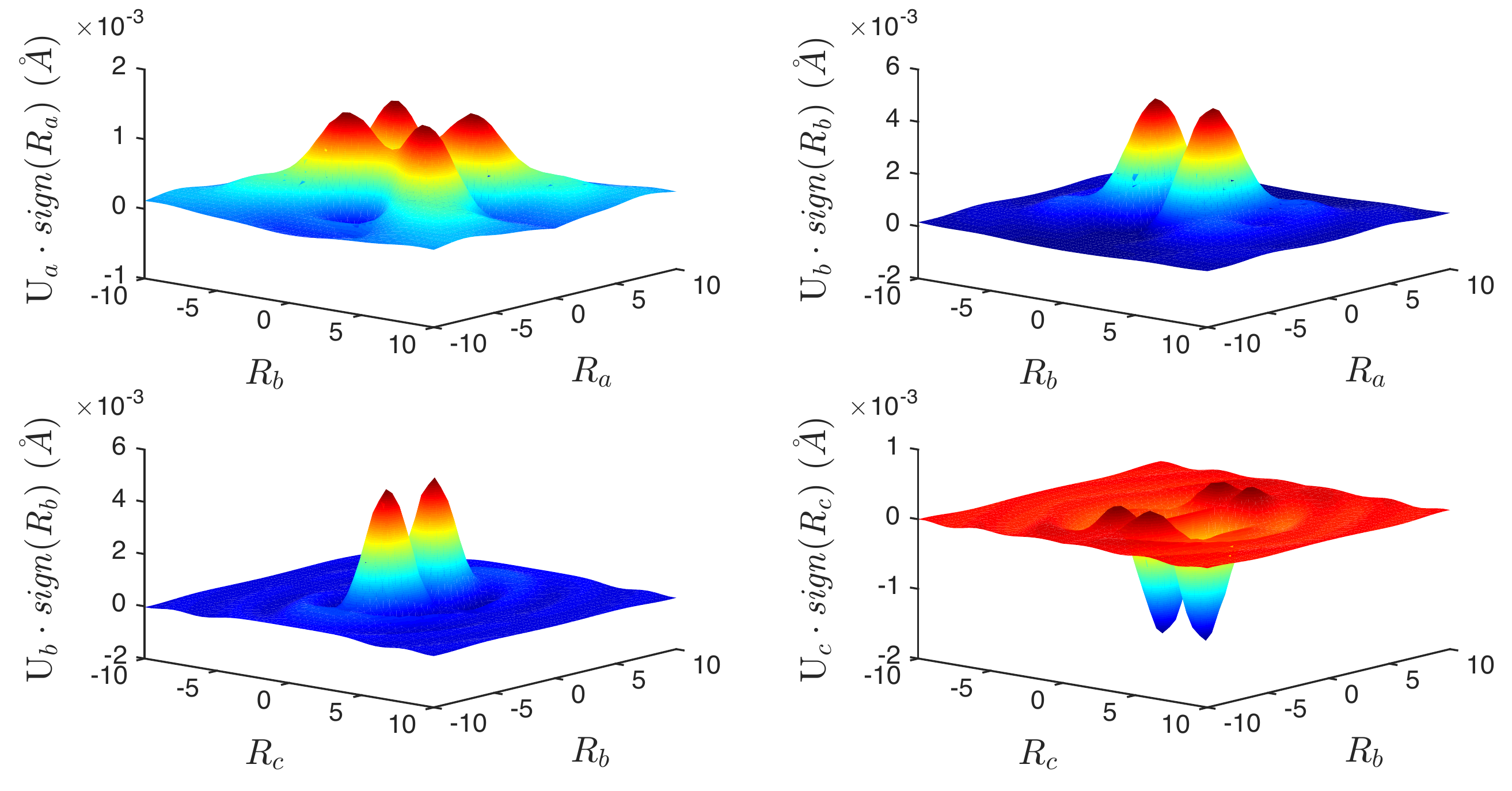}
\caption{Reconstructed acoustic type strain deformation pattern in the $[R_a, R_b, 0]$ (top panels) and $[0, R_b, R_c]$ (bottom panels) planes with $[0, 0, 0]$ to be the dopant center. The distance $R$ is given in unit of unit cell size. $U_{a,b,c}$ are the components of the deformation vector $\vec{U}(\vec{R})$.}
\label{pic:disV}
\end{figure}

With the previously determined cut-off, the reconstructed strain field is shown in Fig.~\ref{pic:disV}. The dopants introduce ripple-like strain field around themselves. The maximum deformation is of the order of 0.001$\AA$, and peaks within $\sim 1nm$ region around the dopants. For the unit cells farther away, their deviation from average position quickly drops to be an-order-of-magnitude smaller. Although the employed cut-off for the integration in Eqn.(\ref{eqn:displacement}) introduces some ambiguity to the reconstructed strain field, we argue that the estimated amplitude is reasonable qualitatively. Comparing with the thermal factors estimated for La$_2$CuO$_4$ at low temperature by neutron powder refinement\cite{Chaillout1990}, our estimated amplitude of the strain field is slightly smaller on average. This is in good agreement with our observations. In Fig.(\ref{pic:qRelation})(b-e), the fitting to several cuts in the reciprocal space, together with the TDS contributions, are shown. At low temperature of 7 K, the TDS contribution to the diffuse scattering intensity is comparable to that from HDS in small $q$ region, and even dominates large $q$ region since the HDS drops faster as function of increasing $|q|$. 

To evaluate the impact of this dopant associated strain field on the local electronic properties in La$_{2-x}$Sr$_x$CuO$_4$, we compare it to the uniform strain effect observed in the pressure experiments\cite{hydrostatic,uniaxial}. In the hydrostatic and uniaxial pressure experiments, the superconducting temperature ($T_c$) for La$_{2-x}$Sr$_x$CuO$_4$ can be tuned of the order of 1K/GPa. Based on the linear compressibility coefficients for La$_{2-x}$Sr$_x$CuO$_4$ measured by G. Oomi et al.\cite{Oomi1991}, a 0.01$\AA$ change in the unit cell size is equivalent to a local pressure of about 1 GPa. Thus, the strain field in La$_{2-x}$Sr$_x$CuO$_4$ we observed is likely to be too small to be responsible for the remarkable nanoscale electronic inhomogeneity observed with STM on La$_{2-x}$Sr$_x$CuO$_4$\cite{Kato1,Kato2}. We noticed that Fujita {\it et al.} reported a dramatic effect of lattice deformation on $T_c$ in Bi$_2$Sr$_{1.6}$Ln$_{0.4}$CuO$_{6+\delta}$ by changing the rare earth ions Ln with different ionic radius\cite{Fujita}. Since the propagated nanoscale strain is small, the leading variables for such effect are likely to be intra-unit cell parameters, such as the local tilting of the CuO$_6$ octahedra or apical oxygen motion, as suggested by the EXAFS experiements\cite{EXAFS1, EXAFS2, EXAFS3}.          

Similar X-ray measurements were carried out by E. D. Isaacs {\it et al}. on La$_{2-x}$Sr$_x$CuO$_4$ with similar doping\cite{Isaacs}. There the diffuse scattering pattern was discussed in the context of lattice response to possible correlated fluctuations of the doped holes. We show that, with the elastic continuum approximation, the scattering from the strain field associated with un-correlated defect centers can well reproduce the measured data. The diffuse scattering peaks discussed in \cite{Isaacs} are from the intensity ridges shown in Fig.(\ref{fig:HKLmesh})(a-c), which are characteristic HDS signals.     

In conclusion, the nanoscale strain associated with the Sr dopants in La$_{2-x}$Sr$_x$CuO$_4$($x=0.08$) has been studied with X-ray diffuse scattering. The observed diffuse scattering pattern can be well modeled with the elastic continuum approximation. From qualitative analysis, we show that the amplitude of this strain field is comparable or smaller than the thermal vibrations at even 7 K. With the lattice deformation being of the order of 0.001 $\AA$ within $\sim{1}${\it nm} around the dopants and drops rapidly farther away, the impact to the local electronic structure from such nanoscale strain field is likely to be negligible. As doping is a routine means for tuning properties of materials, we expect that such HDS analysis presented herein to be a powerful tool to isolate dopant-induced lattice strain and determine its role in electronic properties in a diversity of materials. Furthermore, with coherent-x-ray beam in future diffraction-limited synchrotron radiation sources such diffuse-scattering studies may allow one to explore dynamic fluctuations as well.

We thank J. Tranquada for fruitful discussions. The work at the Institute of Physics, Chinese Academy of Sciences is supported by MOST (Grant No.2015CB921302), CAS (Grant No. XDB07020200). This research used Sector 4 of the Advanced Photon Source, a U.S.Department of Energy (DOE) Office of Science User Facility operated for the DOE Office of Science by Argonne National Laboratory under Contract No. DE-AC02-06CH11357.

\end{document}